\documentclass[]{article}
\usepackage[utf8]{inputenc}
\usepackage{authblk}
\usepackage{graphicx}
\usepackage{amsmath}
\usepackage{makecell}
\usepackage{commath}
\usepackage{natbib}
\usepackage{tabularx}

\begin{document}
\title{A Rich Source of Labels for Deep Network Models of the Primate Dorsal Visual Stream}
\author[1]{Omid Rezai}
\author[2]{Pinar Boyraz Jentsch}
\author[1]{Bryan Tripp}
\affil[1]{Systems Design Engineering \& Centre for Theoretical Neuroscience, University of Waterloo, Waterloo, Ontario, Canada}
\affil[2]{BAST Inc., Leicestershire, UK}
\date{\vspace{-5ex}}
\maketitle

\begin{abstract}
Deep convolutional neural networks (CNNs) have structures that are loosely related to that of the primate visual cortex. Surprisingly, when these networks are trained for object classification, the activity of their early, intermediate, and later layers becomes closely related to activity patterns in corresponding parts of the primate ventral visual stream. The activity statistics are far from identical, but perhaps remaining differences can be minimized in order to produce artificial networks with highly brain-like activity and performance, which would provide a rich source of insight into primate vision. One way to align CNN activity more closely with neural activity is to add cost functions that directly drive deep layers to approximate neural recordings. However, suitably large datasets are particularly difficult to obtain for deep structures, such as the primate middle temporal area (MT). To work around this barrier, we have developed a rich empirical model of activity in MT. The model is pixel-computable, so it can provide an arbitrarily large (but approximate) set of labels to better guide learning in the corresponding layers of deep networks. Our model approximates a number of MT phenomena more closely than previous models. Furthermore, our model approximates population statistics in detail through fourteen parameter distributions that we estimated from the electrophysiology literature. In general, deep networks with internal representations that closely approximate those of the brain may help to clarify the mechanisms that produce these representations, and the roles of various properties of these representations in performance of vision tasks. Although our empirical model inevitably differs from real neural activity, it allows tuning properties to be modulated independently, which may allow very detailed exploration of the origins and functional roles of these properties. 
\end{abstract}


\section{Introduction}
Among primates, the visual system of the rhesus macaque, an old-world monkey with a visual system similar to that of humans, has been studied in the most detail. The macaque visual cortex consists of about thirty distinct visual regions \citep{Felleman1991} that make up a richly connected network \citep{Markov2014}. Responses to visual stimuli have been studied extensively in many of these regions \citep{Orban2008,Kruger2013}. 



These networks are intricate, with hundreds of millions of neurons and multiple levels of organization. Individual neurons are also quite complex \citep[e.g.][]{Markram2015}, with a wide variety of ion channels, dynamics on a wide range of time scales \citep{lundstrom2008fractional}, complex and electrically significant dendrite morphologies \citep{schaefer2003coincidence}, and nonlinear interactions within dendrites \citep{Polsky2004}. Nonetheless, highly abstract deep neural networks not only perform closely related functions \citep{He2015}, but also exhibit closely related internal activity \citep{Yamins2014,Khaligh-Razavi2014,Cadieu2014,Hong2016}. 
Thus highly abstract models may be sufficient to explain human-level core object recognition (a major function of the visual cortex) and many details of the associated neural representations. 


\subsection{Activity in CNNs and the Visual Cortex}
Relationships between activity in the primate visual cortex and CNNs has been studied mostly in the context of the ventral visual stream, which is involved in object recognition and classification \citep{dicarlo2012does}. In CNNs that are trained for object classification in natural scenes, early layers often have Gabor-like receptive fields \citep{Krizhevsky2012} that resemble the receptive fields of neurons in the primary visual cortical area (V1). Activity in later layers of the CNNs is relatively invariant to changes in position, size, and orientation \citep{Zeiler2014}, as is activity in higher visual areas in the brain. Furthermore, activity in intermediate and later layers, respectively, is closely related to activity of real neurons in V4 and the inferotemporal cortex (IT), which occupy similar positions in the ventral visual stream. Specifically, multilinear regression with CNN unit activities accounts well the activity of real neurons \citep{Yamins2014,Cadieu2014}. Suitable combinations of the CNN unit activity are also correlated within object categories, in much the same way as IT neuron activity \citep{Yamins2014}, and to a greater extent that previous ventral-stream models \citep{Khaligh-Razavi2014}. Later CNN layers and IT neurons also contain similar information about non-category stimulus properties such as object position \citep{Hong2016}.

CNN units themselves (as opposed to optimal linear combinations of these units) have different correlation patterns than IT neurons \citep{Yamins2014,Khaligh-Razavi2014}. Relatedly, we found in unpublished work that certain response properties of CNN units (e.g. the distribution of size-tuning bandwidths) closely resembled those in IT, while others (e.g. orientation tuning curves) were quite different. This suggests that there is room to further align CNNs with the ventral stream. There has been little comparison of the dorsal visual stream (which processes spatial and action-related visual information) with CNNs that are trained for spatial tasks. However, \citet{Guclu2016} found that networks trained for action recognition could predict functional magnetic resonance imaging (fMRI) data from the dorsal stream. 

To develop CNNs that more closely resemble the primate visual system, it may be necessary to incorporate additional brain-like mechanisms \citep[e.g.][]{Rubin2015}. Another complementary approach is to directly train deep layers to approximate some kind of neural data. \citet{Yamins2014} tried this, and found that it did not lead to better approximations of a validation dataset of neural activity than simply training the network for object classification. However, their neural dataset was small (5760 images). Much larger datasets are becoming available \citep{minderer2012chronic,obien2015revealing}. However, unless large parts of the brain are removed, these methods are so far limited to superficial areas, whereas much of the macaque visual cortex is within sulci. Small arrays of up to 11 electrodes have been driven together for parallel recordings deep in primate sulci \citep{Diogo2003}, but these have been small-scale acute recordings, whereas large-scale chronic recordings are needed to provide many training examples. 


\subsection{Empirical Activity Models}
An alternative to using large-scale chronic recordings as a source of deep-layer labels is to generate an approximate set of labels for a given visual area, using an empirical model derived from the electrophysiology literature \citep{Tripp2016}. This may be the only way to generate large neural datasets in the near future for areas within cortical sulci. This approach also has an advantage over using real data, in that a model allows artificial variation of tuning properties. For example, one could modulate tuning curve widths, or correlations between different parameters, and explore both the ability of CNNs to reproduce these variations, and their effects on performance of visual tasks. Furthermore, an empirical model may be better suited for approximating the effects of attention on neural activity, because the model can produce activity that is consistent with the network's focus of attention (rather than the animal's focus).

To explore this approach, we have developed a sophisticated model of activity in the primate middle temporal area (MT), an area deep in the superior temporal sulcus that is important for processing of visual motion. Below we briefly review MT physiology. 

\subsection{Area MT}
The middle temporal cortex (MT) receives strong feedforward input from early visual areas V1, V2, and V3 \citep{Maunsell1983,Markov2014}, as well as direct sub-cortical input \citep{sincich2004bypassing,Born2005}. It projects to the higher-level middle superior temporal and ventral intraparietal areas, and also receives strong feedback connections from these areas. Electrical stimulation of MT affects perception of visual motion \citep{Nichols2002}. Inactivation or damage of MT impairs motion perception \citep{newsome1988selective,rudolph1999transient} and the ability to smoothly follow a moving object with the eyes \citep{newsome1985deficits}. Illusions in speed perception have also been linked with subtle properties of MT neuron responses \citep{Boyraz2011}.

Consistent with these effects, many neurons in MT respond strongly to visual motion. 
The spike rates of individual MT neurons vary with a number of stimulus features, including direction and speed of visual motion, and binocular disparity. Many MT neurons are sensitive to motion in depth, i.e. toward or away from the eyes \citep{Czuba2014}. MT is the earliest visual region in which a substantial number of neurons solve the motion ``aperture problem'', responding to the actual direction of motion of a stimulus, rather than the component of motion that is orthogonal to local edges \citep{Pack2001,Smith2005}, which requires only local computations. 

In summary, MT exhibits a particular representation of visual motion, which is similar in scope to scene flow \citep{Mayer2015}. We are interested in whether the particular representation of this information in MT, one of the most thoroughly studied visual areas, can be precisely reproduced in deep artificial neural networks. We have developed the present model to provide synthetic datasets to help address this question. 


\section{Methods}
See (https://github.com/omidsrezai/empirical-mt) for model code. 

\subsection{Model Structure}
Our model is pixel-computable, i.e. it produces approximations of MT spike rates directly from input images. We focus on producing spike rates, rather than spike sequences. As an aside, given these rates, it is straightforward to produce Poisson spike sequences \citep{Dayan2001}, including those with noise correlations that are realistic for MT \citep{Tripp2012}. 

\begin{figure}
\centering
\includegraphics[width=3in]{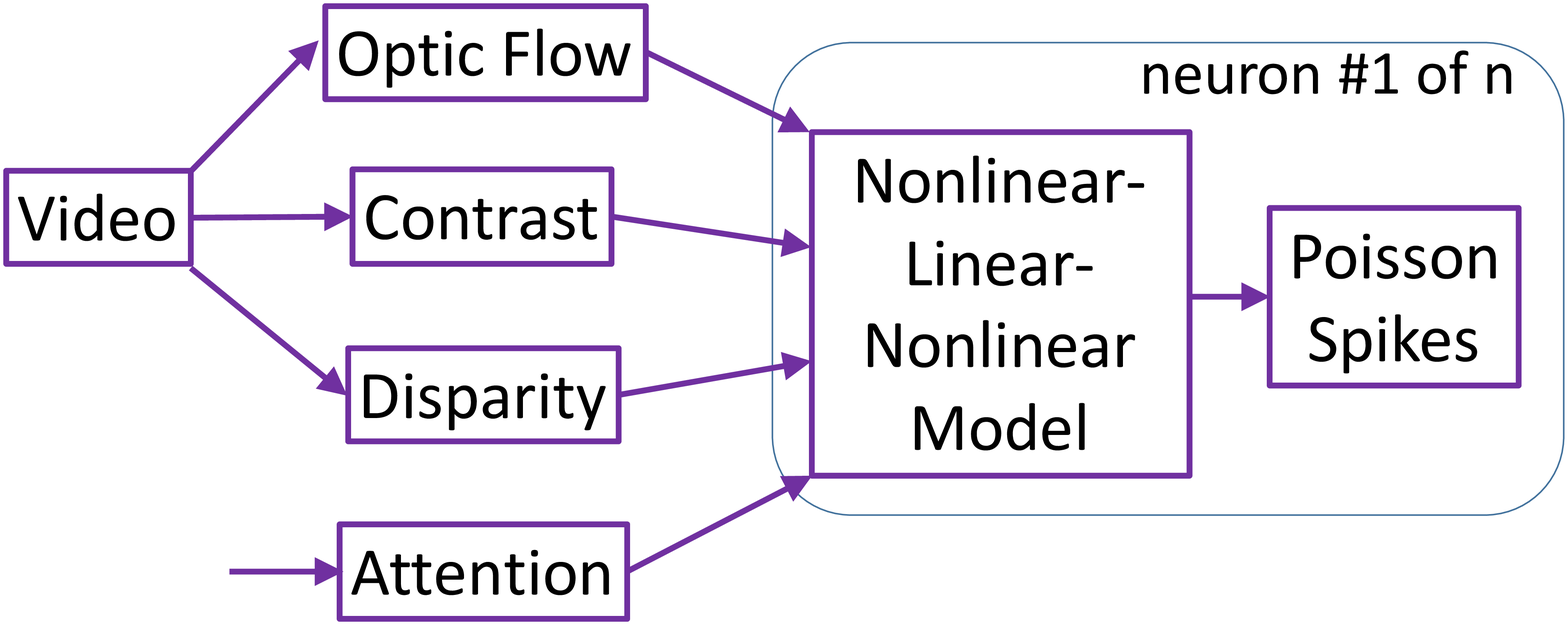}
\caption{Model structure. The model uses nonlinear-linear-nonlinear models to approximate neuron responses as functions of optic flow, contrast, disparity, and attention fields. Optic flow, contrast, and disparity are calculated from input images, as described in the text. 
Poisson spikes can optionally be generated at the estimated spike rates to emulate neural activity more closely, but they are not used in this paper.}
\label{fig:model-sketch}
\end{figure}

The model structure is sketched in Figure \ref{fig:model-sketch}. The model requires five fields as input. The field values are defined at each image pixel $x,y$. The five fields are $u(x,y)$ (horizontal flow velocity), $v(x,y)$ (vertical flow velocity), $d(x,y)$ (disparity), $c(x,y)$ (contrast), and $a(x,y)$ (attention). Section \ref{sec:input-fields} below discusses calculation of these fields. 

The response of each neuron is approximated as a nonlinear-linear-nonlinear (NLN) function of these fields. The first nonlinear step requires calculation of four additional fields for each neuron, each of which is a point-wise nonlinear function of the five input fields. We refer to these functions as tuning functions (see details in Section \ref{sec:tuning-functions}). Each of these tuning functions is used to scale the neuron's response to a different stimulus feature. Specifically, we calculate  $g_s(u,v,c)$ (a function of flow speed and contrast), $g_{\theta}(u,v)$ (a function of flow direction), $g_d(d)$ (a function of disparity), and $g_g(a,c)$ (a function of attention and contrast). 

The full model therefore requires calculation of four times as many of these tuning-function fields as there neurons with distinct sets of parameters. Since the model is intended for training convolutional networks, there is one such set of parameters per channel in the MT layer. This number can be specified at run time, but we would expect it to normally be on the order of 500-2000, so something like 2000-8000 of these fields must be calculated by the full model. One additional field per neuron is then calculated as the point-wise product of these fields \citep[consistent with data from][]{rodman1987coding,Treue1999}. We refer to this as the neuron's tuning field, 
\begin{equation}
t(x,y) = g_s g_{\theta} g_d g_g.
\end{equation}
This completes the first nonlinear stage of the NLN model. Since the model is meant to train convolutional networks, only one tuning field is needed per MT channel (corresponding to a set of model parameters), regardless of the pixel dimensions of the channel. Henceforward, when we talk about a ``neuron model'', it should be understood that this ``neuron model'' is ultimately tiled across the MT layer to simulate many neurons with different receptive field centres.

The remaining linear and nonlinear steps consist of a conventional convolutional layer, with one channel per MT neuron. Kernels combine tuning-field values $t(x,y)$ over a receptive field. However (in contrast with typical convolutional layers), kernels are parameterized to resemble MT receptive fields . The kernels include excitatory, direction-selective suppressive, and non-selective suppressive components. Such components have been found to account well for MT responses to complex motion stimuli \citep{Cui2013}. The excitatory component of the kernel models the neuron's classical receptive field. This component has positive weights and a Gaussian structure. It spans a single channel of the tuning-field layer, and therefore has a speed and direction selectivity that match that channel. The direction-selective suppressive component also spans a single tuning-function channel. It has negative weights, and is modelled as a Gaussian function that may be larger than the excitatory kernel, or offset from it, or elongated. For each neuron, we draw at random from these spatial relationships with the proportions reported by \citet{Xiao1997}. The preferred direction of this suppressive component is generally different from that of the excitatory component. We draw this difference from the distribution in \citet{Cui2013} (their Figure 5). Finally, the non-direction-selective suppressive component has uniform weights across all tuning-layer channels. It has negative weights and an annular structure that we model as a rectified difference of Gaussians. The full kernel is the sum of these components. These kernel structures are somewhat oversimplified, and are the subject of ongoing work. Another possible approach for future work would be to hold the rest of our model constant, learn optimal kernels for tasks that are related to the dorsal stream, and assess whether the statistics of the learned kernels are consistent with the MT data.  

When we fit tuning curves for speed, disparity, and direction tuning in response to stimuli that were spatially uniform in these properties, we simplified the kernels as broad Gaussian functions. 

The final nonlinearity is, 
\begin{equation}
f(x) = \left[ Ax + B \right]^n_+,
\label{eq:output-nonlinearity}
\end{equation}
composed of a half-wave rectification ($[]_+$) followed by a power function ($[]^n$). $A$ and $B$ are a scaling factor and a background spike rate, respectively. 

\subsection{Eccentricity and Receptive Field Size}
The visual cortex differs from convolutional networks in that the receptive fields of neurons in many visual areas scale almost linearly with eccentricity (visual angle from the fovea). This difference could be reduced by remapping the input images. However, we instead model the whole visual field uniformly, as is typical in convolutional networks. There is also variation in receptive field sizes at any given eccentricity. We modelled the spread of receptive field sizes on parafoveal receptive fields (2-10 degree eccentricity) from Figure 2 of \citet{maunsell1987}.

\subsection{Input Fields}
\label{sec:input-fields}
The model requires contrast, attention, optic flow, and binocular disparity fields. 

\subsubsection{Contrast Field}
The contrast field is calculated using the definition of \citet{Peli1990}. This is is a local, band-limited measure, in contrast with other notions of contrast (e.g. root-mean-squared luminance) that are global and frequency-independent. A local definition is needed to modulate neuron responses according to contrast within their receptive fields (as opposed to remote parts of the image). Frequency dependence allows us to match the contrast definition to primate contrast sensitivity \citep{robson1966spatial,de1982spatial}.  

In Peli's definition, contrast at each spatial frequency band (i) is defined as a ratio of two functions, 
\begin{equation}
c_i(x,y) = \frac{\alpha_i(x,y)}{l_i(x,y)}. 
\end{equation}
The numerator function is, 
\begin{equation}
\alpha_i(x,y) = I(x,y)*g_i(x,y), 
\end{equation}
where $I$ is the image, $g_i$ is a spatial frequency dependent filter, and $*$ denotes convolution. The denominator function is,  
\begin{equation}
l_i(x,y) = \bar{I} +  \sum_{j=1}^{i-1} \alpha_j(x,y), 
\end{equation}
where $\bar{I}$ is the image mean. Peli suggested cosine log filters as the choice for $g$s since an image filtered by a bank of these filters can be reconstructed by a simple addition process without distortion. However, to relate the contrast definition more directly to V1, we instead used a bank of Gabor filters with four different frequencies and four different orientations (for a total of 16 contrast channels). We combined these channels in a weighted sum, with weights chosen to approximate macaque contrast sensitivity.
We then smoothed the resulting contrast field with a 2D Gaussian kernel ($\sigma=10$pixels) that was meant to approximate integration over V1 cells, and scaled it so that its mean over the image was equal to the root-mean-squared contrast measure. 

\subsubsection{Attention Field}
Attention is typically driven by task demands, so in general it can not be derived from images alone. Recent models approximate top-down influences \citep{Borji2013}. However, in the context of training neural networks that have attention mechanisms \citep[e.g.][]{Xu2015}, the attention field should ideally be defined by the network itself, to align attention modulation of activity with the network's focus of attention. Therefore we treated the attention field as an input to the model. To test the model, and to compare its output with electrophysiology data, we manually defined attended stimulus regions by drawing polygons around them in a custom user interface.

\subsubsection{Flow and Disparity Fields}
Flow and disparity fields were calculated using computer-vision algorithms. 
Specifically, we used the Lucas-Kanade method \citep{Lucas1981} to estimate both optic flow and disparity from images. Interestingly, a variety of other algorithms produced poorer results, including semi-global matching \citep{hirschmuller2005accurate} and loopy belief propagation on a Markov random field \citep{felzenszwalb2006efficient}. These methods extrapolated beyond well-textured regions in ways that were inconsistent with MT activity. Exploring and perhaps rectifying these differences is an interesting topic for future work. However, the Lucas-Kanade method generally produced better fits with the MT data than previous MT models (see Results).

The classical Lucas-Kanade algorithm does not capture large displacements, but this limitation is addressed by a multi-scale version of the algorithm \citep{Marzat2009}. In this version, the Gaussian pyramids method is used to repeatedly halve the image resolution. Flow or disparity is then estimated at the lowest resolution first. Then at each finer resolution, the immediate lower-resolution estimate is used to warp the earlier image, and the Lucas-Kanade algorithm is used to find residual differences between the warped earlier image and the later image. The multi-scale version of the algorithm also helps to solve the aperture problem, since it finds estimates that are consistent with global motion apparent in downsampled images. We typically used the multiscale algorithm in our simulations, with 4-5 scales. To simulate combined local and pattern motion selectivity \citep{Pack2001}, we mixed the outputs of single-scale and multi-scale versions of the algorithm. 

\subsection{Tuning Functions}
\label{sec:tuning-functions}
Given these fields, the next step in approximating a neuron's activity was calculation of a new four-channel image that consisted of pixel-wise nonlinear functions of the fields. Specifically, we calculated  $g_s(u,v,c)$ (a function of flow speed and contrast), $g_{\theta}(u,v)$ (a function of flow direction), $g_d(d)$ (a function of disparity), and $g_g(a,c)$ (a function of attention and contrast). These functions were adopted from previous studies, as described below.  

\subsubsection{Speed Tuning}
We used a contrast-dependent speed tuning function, \citep{Nover2005}, 
\begin{equation}
g_s = \exp \left(- \frac{ [\log \left( q(s,c)\right)]^2}{2\sigma_s^2}\right), \end{equation}
where, 
\begin{equation}
q(s,c) = \frac{s+ s_0}{s_p(c) + s_0},
\end{equation}
$s=\sqrt{u^2+v^2}$ is motion speed, $s_p$ is the preferred speed. The tuning curve has parameters $s_0$ (offset) and $\sigma_s$ (width). Preferred speed is a function of contrast,
\begin{equation}
s_p(c) = \frac{A_p c}{c+B_p},
\label{eq:prfSpCnt}
\end{equation} where $A_p$ and $B_p$ are additional parameters that define a saturating dependence of preferred speed on contrast. 

Note that many MT neurons are selective for spatiotemporal frequency, rather than speed, when stimulated with sinusoidal gratings \citep{priebe2006tuning}. However more complex stimuli that contain many frequencies elicit responses to motion speed. 

\subsubsection{Direction Tuning}
Direction tuning was modeled as \citep{Wang2016},
\begin{equation}
g_{\theta} = \exp \left( \frac{cos \left(\theta - \theta_p\right)-1 }{\sigma_{\theta}}\right)+ a_n \exp \left( \frac{cos \left(\theta - \theta_p - \pi \right)-1 }{\sigma_{\theta}}\right),
\end{equation}
where $\theta= atan2(v,u)$ is motion direction, $\theta_p$, $\sigma_{\theta}$, and $a_n$ are the preferred direction, direction width, and relative amplitude in null direction (i.e. 180 degrees away from preferred direction), respectively. 

\subsubsection{Disparity Tuning}
Similarly, disparity tuning was modeled using Gabor functions \citep{DeAngelis2003}, 
\begin{equation}
g_d = \exp \left( \frac{-\left(d - d_p \right)^2}{2\sigma_d^2}\right) \cos \left( 2 \pi f_d (d - d_p) +\phi_d \right),
\end{equation}
where $d_p$ and $\sigma_d$ set the centre and width of the Gaussian component and $f_d$ and $\phi_d$ are the frequency and phase of the oscillatory component.

\subsubsection{Attention}
Lastly, the gain function was \citep{Treue1999,martinez2002},
\begin{equation}
g_g(a,c) = \begin{cases}
A_g g_c(c), & \text{if } a = 1\\
\\
g_c(c), & \text{if } a=0\\  
\end{cases}
\end{equation}
where $A_g$ is the attentional gain and $g_c$, is the contrast response function defined as:
\begin{equation}
g_c(c) = \frac{A_c c^n_c}{c^n_c+B_c}.
\end{equation}


\subsection{Extension to a Binocular Model}
In many of the electrophysiology experiments that inform the model, monkeys were free to converge their eyes on a single, flat computer display, bringing binocular disparity close to zero. Recent evidence indicates that in the presence of stereo stimuli, some MT neurons are tuned for motion-in-depth \citep{Czuba2014}. To account for such 3D motion encoding of MT neurons, we extended our model to a binocular version by modifying Equation \ref{eq:output-nonlinearity} as,
\begin{equation}
f(x) = \left[ A_L x_L + A_R x_R + B \right]^n_+,
\label{eq:output-nonlinearityBIN}
\end{equation}
where $A_L$ and $A_R$ are left and right eye gains, and $x_L$ and $x_R$ are weighted sums of tuning functions in left and right eye respectively.

\subsection{Tuning Curve Fits}
To test the model, we fit various tuning curves from the electrophysiology literature using Matlab's nonlinear least-squares curve fitting function, \emph{lsqcurvefit}. The fitting procedure for a given tuning curve selected the parameters of the relevant tuning functions (e.g. $g_s(u,v,c)$), along with parameters $A$ and $B$ of Equation \ref{eq:output-nonlinearity}. As the optimization was non-convex, we initiated it from at least 100 different starting points for each neuron, and took the best answer. 

This approach was designed to have a high success rate, in order to reliably support development of a rich statistical model of MT activity. Aside from failures of the optimization procedure (which we minimized by restarting from many initial parameter values), the approach has two potential failure modes. The first would arise from a poor choice of nonlinear function, however we chose functions that are well supported by previous work. The second would be a failure of the computer vision algorithms to estimate the relevant parameters from the images. We generally had good results with the Lucas-Kanade algorithm, although several other computer vision algorithms (including more sophisticated ones) produced poorer matches. 


\subsection{Parameter Distributions}
For simulations with many neurons, we drew the neurons' tuning parameters from statistical distributions that were based on histograms and scatterplots in various MT electrophysiology papers. The model required distributions of preferred disparity, preferred speed, speed tuning width, 
attentional index \citep{Treue1999}, 
and a number of other tuning properties. As a first step in approximating these distributions, we extracted histograms and scatterplots of various tuning properties from the literature using Web Plot
Digitizer (http://arohatgi.info/WebPlotDigitizer/). We then modeled each histogram using either a standard distribution (one of Gaussian, log-Gaussian, Gaussian mixture, gamma, t location-scale, exponential, and uniform), or the Parzen-window method \citep{Parzen1962}. For Parzen-window method, we selected the bandwidths using Silverman’s rule of thumb \citep{silverman1986}. In each case, we chose the distribution model that minimized the Akaike Information Criterion \citep{Akaike1974}. The parameter distributions are summarized in Table \ref{tab:param-distributions}. 

\subsection{Correlation between Model Parameters}
To make our model more realistic, we looked for studies that examined the  correlation between the tuning parameters in area MT. \citet{Bradley1998} found that the center-surround effects of disparity and
direction are mainly independent of each other, supporting the way we combine them over the MT receptive field. In another study, \citet{DeAngelis2003} did not find a correlation between direction and disparity tuning parameters. They reported a non-zero correlation between speed and disparity tuning (neurons with higher speed preference tend to have weak and broad disparity tuning). However, this correlation was weak (see their Figure 11.A) and therefore we ignored it in our model. The other correlation they found was between the preferred disparity and the disparity phase of the neurons whose preferred disparity is close to zero. 
We included this correlation by modelling the conditional distribution of disparity phase given the preferred disparity. 

\renewcommand{\arraystretch}{1.5}
\begin{table}
\centering
\caption{Distribution families used for various tuning parameters, and sources in the literature from which distributions were estimated. The number in the bracket specifies the dimension of a parameter, for those that have more than one. 
}
\begin{tabularx}{\textwidth}{c c c}
Parameter & Distribution & Source \\
\hline
Preferred direction & Uniform & \makecell{\citet{DeAngelis2003}} \\
\hline
Direction bandwidth & Gamma  &  \makecell{\citet{Wang2016}} \\
\hline
\makecell{Null-direction \\ amplitude} & t location-scale  & \makecell{\citet{Maunsell1983}}\\
\hline
Preferred speed & Log uniform & \makecell{\citet{Nover2005}}\\
\hline
Speed width & Gamma & \makecell{\citet{Nover2005}} \\
\hline
Speed offset & Gamma & \makecell{\citet{Nover2005}} \\
\hline
Attentional index & t location-scale & \makecell{\citet{Treue1999}}\\
\hline
\makecell{Contrast influence \\ on preferred speed (2)} & 2D Gaussian mixture & \makecell{\citet{pack2005contrast}} \\
\hline
\makecell{Contrast influence \\ on gain (3)} & \makecell{Conditional on \\ attentional index} & \makecell{\citet{martinez2002}} \\
\hline
Preferred disparity & t location-scale & \makecell{\citet{DeAngelis2003}} \\
\hline
Disparity frequency & Log normal &  \makecell{\citet{DeAngelis2003}}  \\
\hline
Disparity phase & \makecell{Gaussian mixture \\ (two components)} &  \makecell{\citet{DeAngelis2003}} \\
\hline
Ocular dominance & t location-scale &  \makecell{\citet{DeAngelis2003}} \\
\hline
CRF size & \makecell {t location-scale} & \makecell{\citet{maunsell1987}} \\
\hline
\end{tabularx}
\label{tab:param-distributions}
\end{table}

\subsection{Comparison with Previous Models}
We compared our model to the models of \citet{Nishimoto2011} and \citet{Baker2016}. We chose these models because they are recent and pixel-computable. Both build on a previous influential MT model \citep{Rust2006}. 

\subsubsection{Nishimoto \& Gallant}
Here we briefly describe the model; see \citet{Nishimoto2011} for more details. Briefly, a video sequence first passes through a large bank of V1-like spatiotemporal filters with rectifying nonlinearities. The filter outputs are combined over local neighborhoods through divisive normalization. Finally, the normalized outputs are weighted optimally to approximate neural data. 

In more detail, a luminance movie $I(x,y,t)$ (where $x$ and $y$ are pixel coordinates and $t$ is time) is first passed through spatiotemporal Gabor kernels to produce linear responses,
\begin{equation}
L_i(t) = \sum_x \sum_y \sum_{\tau} G_i(x,y,\tau) I(x,y,t-\tau), 
\end{equation}
where $i$ is an index, and $G_i$ is the $i^{th}$ Gabor function. Each Gabor function has a certain centre, spatial frequency, phase, and orientation, and temporal frequency. 

V1 simple-cell responses are approximated as, 
\begin{equation}
S_i(t) = \left[ L_i(t) \right]_+^a, 
\end{equation}
where $[]_+$ indicates half-wave rectification and $a$ is an exponent that allows linear, sublinear, and supralinear functions. Their results were best with $a=0.5$. Complex-cell responses $C_j(t)$ were produced as the root-sum-squared output of a quadrature pair of simple cells (i.e. two cells with the same parameters except for a spatial phase difference of $90^{\circ}$). 

Divisive normalization \citep{Carandini2011} was implemented as, 
\begin{equation}
X_j(t) = \frac{C_j(t)}{\sum_j C_j(t) + \beta}, 
\end{equation}
where $\beta$ controls saturation and prevents division by zero. Finally, a spike rate was produced as a weighted sum of normalized responses, 
\begin{equation}
r(t) = \sum_j \sum_\tau w_j(\tau) X_j(t-\tau), 
\end{equation}
where $w_j$ is a weight and $\tau$ is a time offset. The weights were optimized separately for each model neuron. As in \citet{Nishimoto2011}, we used a bank of $N=1296$ filters, including those with spatial frequencies up to two cycles per receptive field. 
 
We used multivariate linear regression to optimize the weights, as in \citet{Rust2006}. 
More specifically, to find the optimized weights, we generated training and testing movies for each tuning curve. Each movie was $2000\times M$ frames in length, where $M$ was the number of data points in the tuning curve. We used the training movie as input to the model and found the weights that minimized the error function, 
\begin{equation}
E(\mathbf{w}) = \norm{\mathbf{X}_{train}\mathbf{w}-\mathbf{R}}^2 +\lambda \norm{\mathbf{{w}}}^2,
\label{eq:error}
\end{equation}
where $\mathbf{w} \in \rm I\!R^{10N}$ is the weight vector, $\mathbf{X}_{train} \in \rm I\!R^{2000M \times 10N}$ is a matrix containing normalized V1 responses when the training movie was used as input, $\mathbf{R} \in \rm I\!R^{2000M}$ is a vector containing the MT responses, and $\lambda$ is the tolerance constant. The optimal weights that minimize this error function can be computed from,
\begin{equation}
 \mathbf{w} = \left(\mathbf{X}_{train}^{\mathit{T}}\mathbf{X}_{train} +\lambda \mathit{I} \right)^{-1}\mathbf{X}_{train}^{\mathit{T}}\mathbf{R},
\end{equation}
where $\mathit{T}$ denotes the transpose, $-1$ denotes the inverse, and $\mathit{I}$ denotes the identity matrix.

To avoid overfitting, we used the testing movie as input and defined $\lambda = 2^p$. Then the value for $p$ was chosen such that,
\begin{equation}
E(\mathbf{w}) = \norm{\mathbf{X}_{test}\mathbf{w}-\mathbf{R}}^2,
\end{equation}
was minimized.

\subsubsection{Baker \& Bair}
The model of \citet{Baker2016} is composed of two cascaded circuits. The first circuit calculates the motion response while the second calculates disparity. However, they used only the first circuit to approximate the motion tuning of MT neurons. We implemented their motion circuity, which is similar to that of Nishimoto \& Gallant, but it includes an additional V1 opponency stage. 

The motion circuity described by \citet{Baker2016} included a population of units tuned to different motion directions. However, their population did not span multiple motion speeds or texture frequencies. To ensure that the model responded realistically to a wider range of stimuli, we replaced their groups of twelve direction-selective units with the filter bank (1296 filters) that we used for the \citet{Nishimoto2011} model. 
We used the same procedure to find the optimal weights as we did for \citet{Nishimoto2011} model, except that due to a nonlinearity at the output of Baker \& Bair's model, we first transformed the tuning curves that we wanted to approximate by the inverse of this nonlinearity. 

\section{Results}

\subsection{Tuning Curve Approximations}
We tested the accuracy with which our model could reproduce tuning curves of real MT neurons from the electrophysiology literature. 
For each tuning curve, we generated the same kinds of visual stimuli (e.g. drifting gratings, plaids, and fields of moving random dots) that were shown to the monkeys. We used these stimuli as input to the model, and optimized the model parameters to best fit the neural data. 

Table \ref{tab:error-report} summarizes the results of the tuning curve fits for our model, which we call Lucas-Kanade Nonlinear-Linear-Nonlinear (LKNLN), and our implementations of the previous models by \citet{Nishimoto2011} (NG) and \citep{Baker2016} (BB). Note that \citet{Baker2016} provide a software implementation of their model, but it has a small filter bank (see Methods) that is inadequate for processing many stimuli. We optimized relevant model parameters individually for each tuning curve. In the LKNLN model, there were relatively few such parameters, because the tuning curves are independent, and we did not change the calculation of the input fields, so only the parameters of the relevant tuning function and final nonlinearity were optimized. For the NG and BB models we optimized all the models' variable parameters, including weights of the spatiotemporal filters, for each tuning curve. Examples of tuning curve fits are shown in the following figures.


\begin{table}
\centering
\begin{tabular}{c|c|c|c|c}
 & \#Tuning Curves & LKNLN & NG & BB \\
\hline
Speed &11 (8) & 0.0531 & 0.1075& 0.1654 \\
\hline
Speed/Contrast & 2 (8) & 0.0543 & 0.2959 &  0.3650 \\
\hline
Attention/Direction &2 (12) & 0.0384  & 0.0848 &  0.1100 \\
\hline
Local motion & 6 (12)& 0.1187 & NA & NA  \\
\hline
Disparity &13 (9) & 0.0316 & NA & -\\
\hline
3D Motion & 8 (12) &0.2144 & NA & 0.2450\\
\hline
Stimulus size & 2 (7) &0.0667 & 0.0841 & 0.0599\\
\end{tabular}
\caption{Summary of RMSE comparison between our model (LKNLN),  \citet{Nishimoto2011} (NG), and \citet{Baker2016} (BB) to the neural data for different tuning parameters. The second column provides the number of tuning curves (along with the number of points in each tuning curve). ``NA'' indicates properties that models do not reproduce according to the literature. For example, the NG model is monocular, so it does not reproduce binocular phenomena. The dash indicates the simulation of the BB model for disparity, which we have not yet performed (we expect BB to perform similarly to our model). 
}
\label{tab:error-report}
\end{table}

\begin{figure}
\centering
\includegraphics[width=4in]{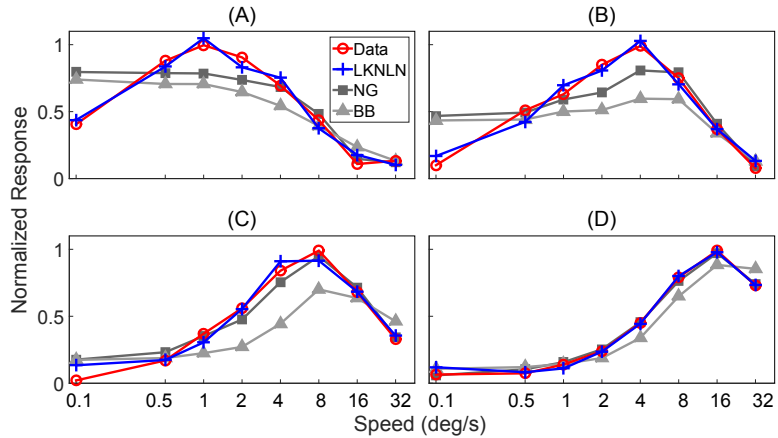}
\caption{Speed tuning curves of four MT neurons, plotted on a logarithmic speed axis. Responses have been normalized so that the peak response of each neuron is equal to 1 . Mean $\pm$ SD error for (A): $0.00\pm 0.06$ spike/s (LKNLN), $0.00\pm 0.18$ spike/s (NG), and $0.06\pm 0.21$ spike/s (BB); for (B): $-0.01\pm 0.06$ spike/s (LKNLN), $-0.00\pm 0.18$ spike/s (NG), and $0.09\pm 0.23$ spike/s (BB); for (C): $-0.01\pm 0.06$ spike/s (LKNLN), $-0.00\pm0.08$ spike/s (NG), and $0.11\pm 0.21$ spike/s (BB); for (D): $-0.00\pm 0.02$ spike/s (LKNLN), $-0.00\pm 0.02$ spike/s (NG), and $0.02\pm 0.09$ spike/s (BB). Data replotted from \citet{Nover2005}.}
\label{fig:speed-Nover}
\end{figure}

Figure \ref{fig:speed-Nover} shows the speed tuning curves of four neurons (with different preferred speeds) where the monkeys were shown fields of random dots moving with different speeds. Our model approximates the neural data more closely than the models of \citet{Nishimoto2011} and \citet{Baker2016}. 

\begin{figure}
\centering
\includegraphics[width=4in]{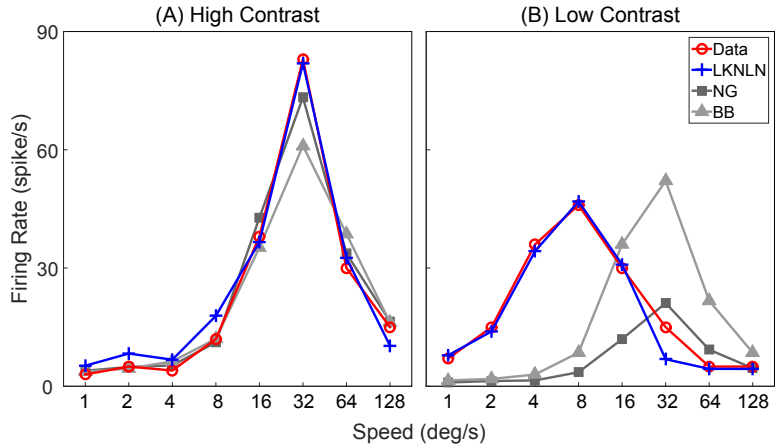}
\caption{Effect of contrast on speed tuning curves. A, Speed tuning in high contrast. Mean $\pm$ SD error: $-1.19\pm 3.35$ spike/s (LKNLN), $ -0.18 \pm 4.40$ spike/s (NG), and $1.55 \pm 8.90$ spike/s (BB). B, Speed tuning in low contrast. Mean $\pm$ SD error: $  1.19 \pm 2.97$ spike/s (LKNLN), $13.09 \pm 17.83$ spike/s (NG), and $3.22 \pm 24.84$ spike/s (BB).
Contrast modulates the response and also shifts the peak (i.e., the preferred speed). While contrast modulates the response amplitude in all three models, only our model (LKNLN) accurately shifts the peak. Data replotted from \citet{pack2005contrast}.}
\label{fig:speed-contrast}
\end{figure}

Figure \ref{fig:speed-contrast} illustrates the speed tuning of a neuron for moving random dots in two cases: when dot luminance was high, resulting a high contrast stimulus (Figure \ref{fig:speed-contrast}A), and when dot luminance was low, resulting a low contrast stimulus (Figure \ref{fig:speed-contrast}B). As shown in the figure, increasing the contrast not only modulated the response gain (peak spike rate) but it also shifted the preferred speed (position of the peak on the speed axis). Our model reproduces both these phenomena, whereas the previous models reproduce only the first. It is important to note that the previous models were intended to provide mechanistic explanations of MT response phenomena, whereas our goal is to imitate MT responses as accurately as possible. Our model does not provide a better explanation for the MT data, only a better fit, which is what is needed to train deep networks. 

\begin{figure}
\centering
\includegraphics[width=4in]{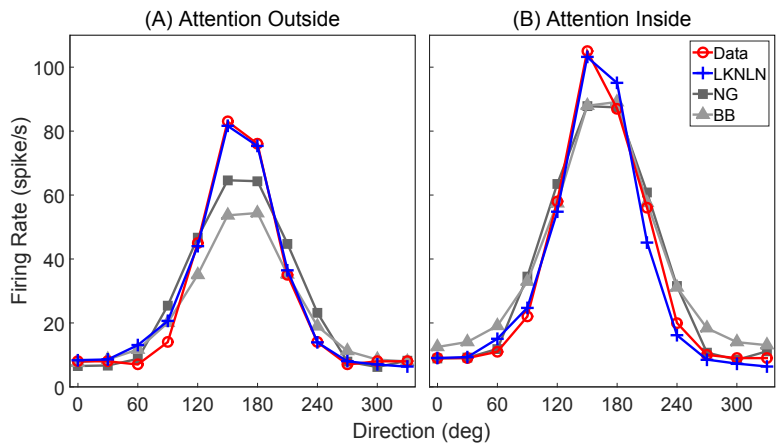}
\caption{Attentional modulation of direction tuning. A, When the stimulus inside the RF was not attended. Mean $\pm$ SD error: $-0.90 \pm 2.73$ spike/s (LKNLN), $-0.02 \pm 8.51$ spike/s (NG), and $3.34 \pm 11.26$ spike/s (BB). B, When the stimulus inside the RF was attended. Mean $\pm$ SD error: $0.90 \pm 4.63$ spike/s (LKNLN), $-1.73 \pm 7.44$ spike/s (NG), and $-3.52 \pm 7.43$ spike/s (BB). Neural data for both cases replotted from \citet{Treue1999}. Our model (i.e., LKNLN) receives attention masks as input, so we defined the masks so that they did not cover the stimulus for the unattended case and covered for the attended case. For the other two models, we first found the best fit for the unattended case by multivariate regression. Given the unattended solution we then found the gain that minimized the error difference between the attended tuning curve and the modulated unattended solution.}
\label{fig:attention}
\end{figure}

Figure \ref{fig:attention} shows our approximation of the effect of attending to stimuli in a neuron's receptive field, which is an increase in gain. The degree of modulation varies across MT neurons. 

\begin{figure}
\centering
\includegraphics[width=3.5in]{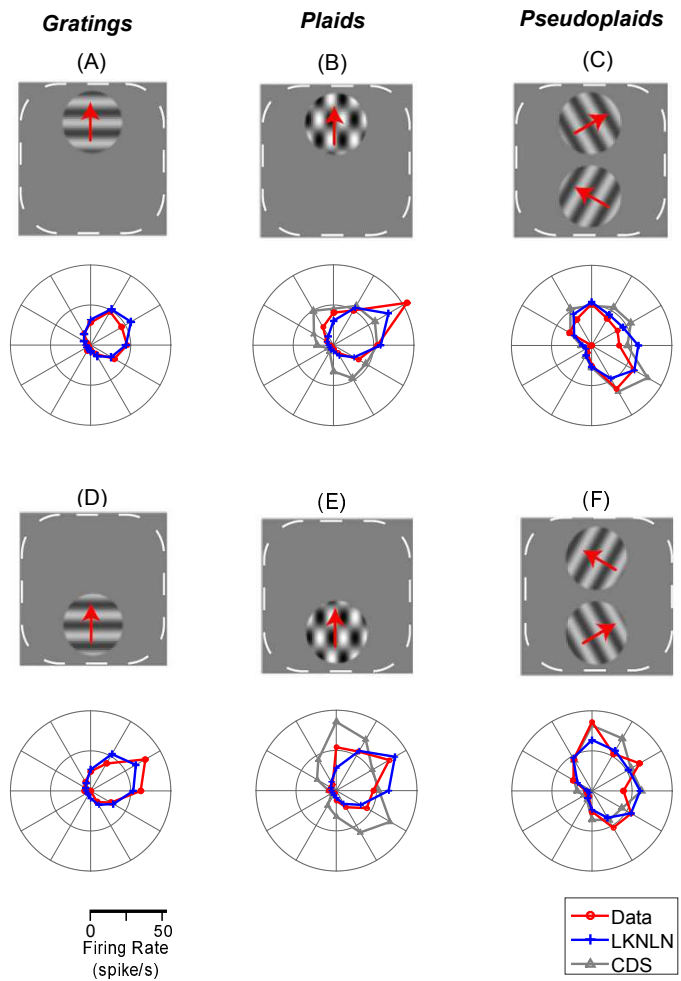}
\caption{Response of an MT cell to gratings and plaids placed within different regions of the cell's receptive field (RF). The response magnitude is plotted on the radial axis, and the angular axis is the direction of motion.
A,D, The neuron's response to grating stimuli at two different patches within RF. B, E, The neuron's response to plaids placed at two different regions over RF. The plaid stimuli are made by overlapping two gratings oriented $120^{\circ}$ apart. Since this cell is selective for the motion of plaids independent of the orientation of their components (gratings), it is classified as a pattern direction selective (PDS) neuron. D, F, The two grating components of the plaids in (B,E) separated to different parts of the receptive field. If motion integration in MT cells were global (i.e., if these cells simply pooled all of their inputs from V1 cells), these plots would be similar plots as (B,E). Instead, the response in this case is close to the component direction selective (CDS) prediction, indicating that motion integration in MT cells are local rather than global. Our model produces realistic responses. Neural data (red) and CDS prediction (gray) replotted from \citet{Majaj2007}; blue is our model.}
\label{fig:local_int}
\end{figure}

\citet{Majaj2007} showed that motion integration by MT neurons occurs locally within small sub-regions of their receptive fields, rather than globally across the full receptive fields. They identified two regions within the receptive fields of a neuron where presenting the stimulus evoked similar neural responses. Then, they studied motion integration by comparing the direction selectivity of MT neurons to overlapping and non-overlapping gratings presented within the receptive field. Since motion integration was local, the ability of the neurons to integrate the motions of the two gratings was compromised when gratings were separated. Our model approximates this neural behavior well (see Figure \ref{fig:local_int}). According to \citet{Nishimoto2011}, their model does not account for this phenomenon, and extending it to do so would require including nonlinear interactions between the V1 filters of the model, which would drastically increase the number of parameters, making estimation more difficult. Other previous models that treat overlapping and non-overlapping features identically \citep{Simoncelli1998,Rust2006,Baker2016} would also not reproduce this phenomenon. 

\begin{figure}
\centering
\includegraphics[width=4in]{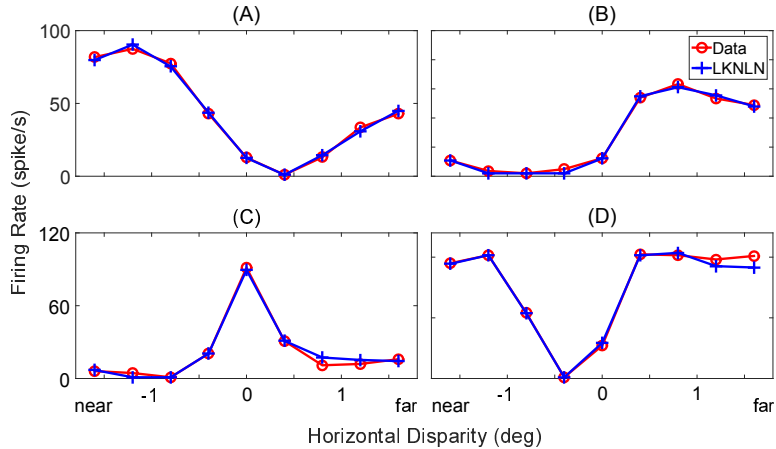}
\caption{Disparity tuning curves of four neurons. Data replotted from \citep{DeAngelis2003}. A, Near ($0.00 \pm 1.96$ spikes/s; mean error $\pm$ SD). B, Far ($0.50 \pm 1.58$ spikes/s; mean error $\pm$ SD). C, Tuned-zero ($-0.43 \pm 2.93$ spikes/s; mean error $\pm$ SD). D, Tuned inhibitory ($1.38 \pm 3.77$ spikes/s; mean error $\pm$ SD). }
\label{fig:disparity}
\end{figure}

MT neurons also encode binocular disparity, with a variety of responses across the MT population, including preferences for near and far disparities, and various selectivities and depths of modulation. 
Our model closely approximates a wide variety of MT neuron disparity-tuning curves (Figure \ref{fig:disparity}). 

\begin{figure}
\centering
\includegraphics[width=4in]{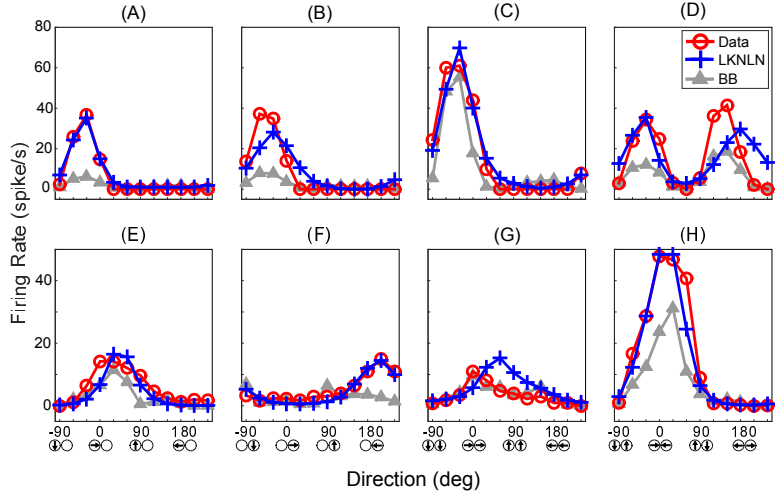}
\caption{Examples of direction tuning of two MT neurons to monocular and binocular stimuli. A–D, An MT neuron tuned for frontoparallel motion. A-B, Direction tuning for gratings presented monocularly to the left (A) and right eye (B). C, Direction tuning for binocular presentation of identical gratings. D, Direction tuning for gratings drifting in opposite directions in the two eyes. E–H, Responses of an MT neuron tuned for motion toward the observer. Direction tuning curves for monocular gratings (E, F), binocularly matched (G), and binocularly opposite motion (H). Neural data replotted from \citet{Czuba2014} in red, prediction of our model (LKNLN) in blue, and prediction of \citet{Baker2016} model (BB) in gray. Mean $\pm$ SD error, A: -1.11 $\pm$ 1.77 spikes/s (LKNLN) and 4.62 $\pm$ 10.63 spikes/s (BB), B: -0.25 $\pm$ 7.04 spikes/s (LKNLN) and 5.70 $\pm$ 11.39 spikes/s (BB), C: -0.61 $\pm$ 5.25 spikes/s (LKNLN) and 5.27 $\pm$ 9.85 spikes/s (BB), D: -0.71 $\pm$ 12.91 spikes/s (LKNLN) and 8.95 $\pm$ 9.40 spikes/s (BB), E: 1.48 $\pm$ 2.88 spikes/s (LKNLN) and 2.84 $\pm$ 3.02 spikes/s (BB), F: 0.52 $\pm$ 1.30 spikes/s (LKNLN) and 2.58 $\pm$ 4.84 spikes/s (BB), G: -2.47$\pm$ 3.98 spikes/s (LKNLN) and -0.45 $\pm$ 1.42 spikes/s (BB), H: 1.42 $\pm$ 4.96 spikes/s, (LKNLN) and 8.15 $\pm$ 10.85 spikes/s (BB). }
\label{fig:3DT}
\end{figure}

Recent studies \citep{Czuba2014} have revealed that some of the MT neurons respond to 3D motion, confirming area MT's role in encoding information about motion in depth, which is vital in accomplishing visually-guided navigation. Figure \ref{fig:3DT} shows the neural responses of two different neurons to monocular and binocular stimuli. One neuron (Figure \ref{fig:3DT}A-D) is tuned for fronto-parallel motion while the other neuron is tuned for motion toward the observer (Figure \ref{fig:3DT}E-H). The binocular version of our model approximates both types of the neurons. 

\begin{figure}
\centering
\includegraphics[width=4in]{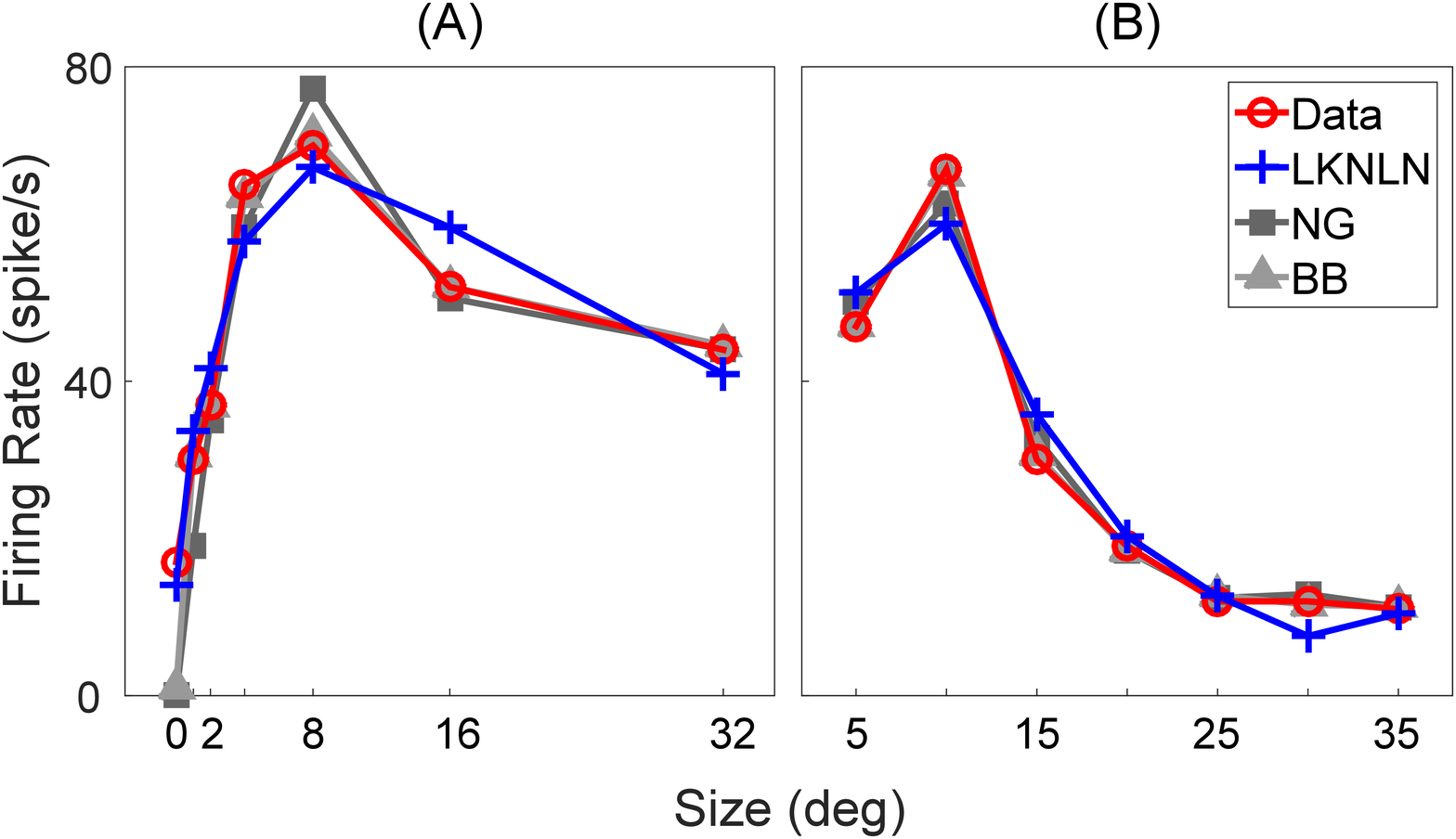}
\caption{Two examples of size tuning curves. The kernels, which gave rise to the size tuning in our model (LKNLN), were radially symmetric difference of Gaussians centered at the center of video frames (the same as neuron's receptive field center). A, Neural data replotted from \citet{DeAngelis2003}. Mean $\pm$ SD error: -0.00 $\pm$ 5.32 spikes/s (LKNLN), 4.21 $\pm$ 7.86 spikes/s (NG), and 2.18 $\pm$ 6.16 spikes/s (BB). B, Neural data replotted from \citet{pack2005contrast}. Mean $\pm$ SD error: 0.00 $\pm$ 4.54 spikes/s (LKNLN), -0.35 $\pm$ 2.50 spikes/s (NG), and -0.01 $\pm$ 0.59 spikes/s (BB).}
\label{fig:sizeTune}
\end{figure}

Size tuning is a result of antagonistic surrounds. Increasing the size of the stimulus to a specific point (optimal size) will increase an MT neuron’s response, while any stimulus larger than the optimal size will often evoke a smaller response. Figure \ref{fig:sizeTune} shows an approximation of two size-tuning curves using a symmetric difference-of-Gaussians kernel, one of three types that we adapt from \citep{Xiao1997}.

\subsection{Parameter Distributions}
Our model is intended to closely approximate population activity in MT. Therefore, statistical distributions of parameters are also an important part of the model. Such distributions have frequently been estimated in the literature. However, past computational models of MT have typically not attempted to produce realistic population responses, except along a small number of tuning dimensions  \citep[e.g.][]{Nover2005}. 

\begin{figure}
\centering
\includegraphics[width=4in]{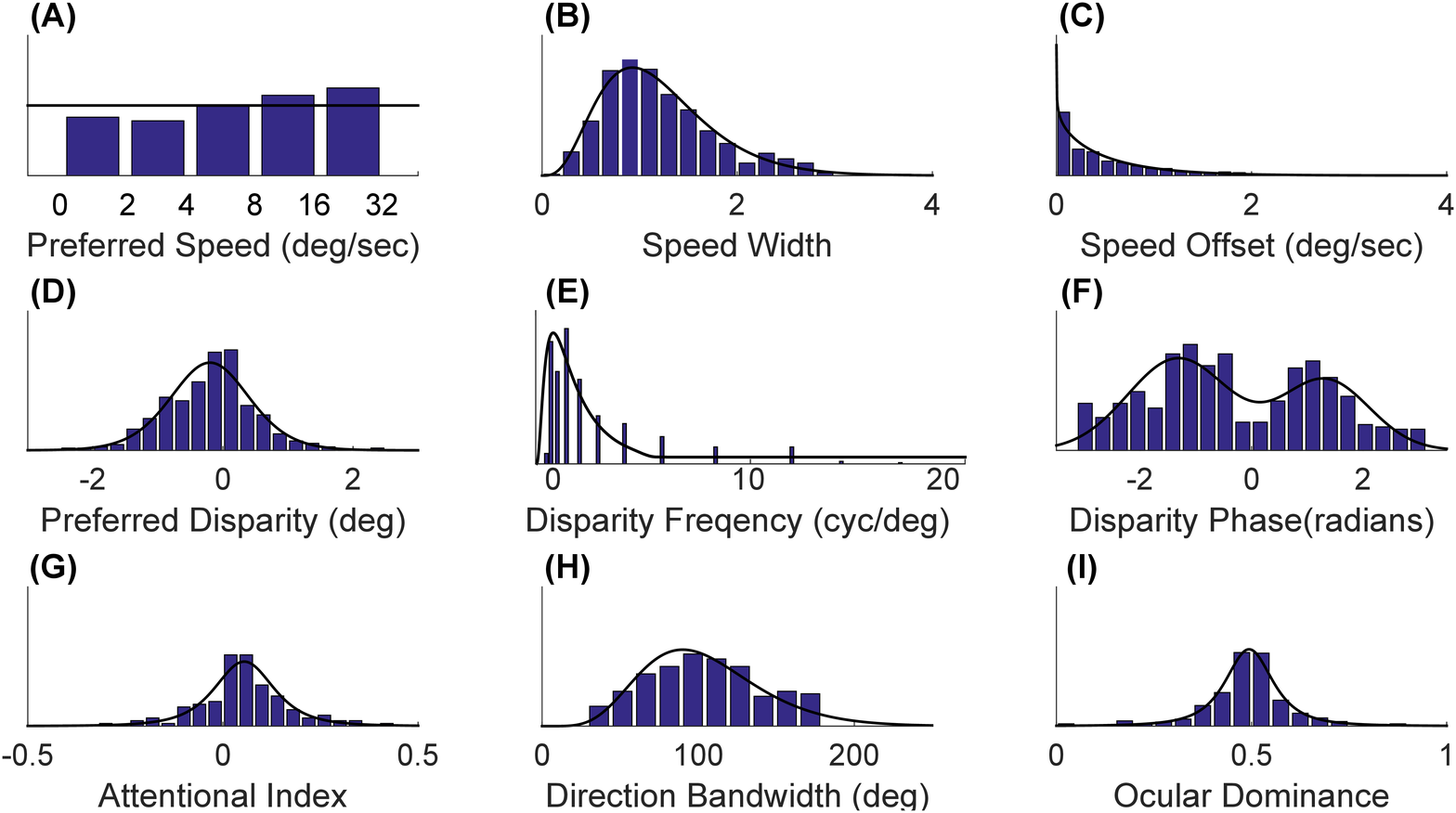}
\caption{Examples of parameter distributions. In each case we replot the data (histograms) along with the selected distribution. A-C, speed parameters including preferred speed (log uniform) in logarithmic space, speed width (gamma), and speed offset (gamma) \citep{Nover2005}. D-F, disparity parameters including preferred disparity (t location-scale), disparity frequency (log normal), and disparity phase (Gaussian mixture) \citep{DeAngelis2003}. G, Attentinal index (t location-scale) \citep{Treue1999}. H, Direction bandwith (gamma) \citep{Wang2016}. I, Ocular dominance (t location-scale) \citep{DeAngelis2003}. }
\label{fig:hist}
\end{figure}

Figure \ref{fig:hist} shows nine examples of fits of parameter distributions. In each case we chose the best of seven different distributions according to the Akaike Information Criterion \citep{Akaike1974}. 

\section{Discussion}
We have developed an artificial, pixel-computable reconstruction of activity in the primate middle temporal visual area (MT) that is more accurate in several ways and more comprehensive than state-of-the art models \citep{Nishimoto2011,Baker2016}. Our model is empirical, so it is not a source of insight into neural mechanisms on its own. However, we think it will provide a powerful way to explore how various aspects of MT-like activity can be produced in deep networks, as well as the roles of its tuning properties in visual tasks (e.g. visual odometry, action recognition, segmentation with motion cues). 

\subsection{Limitations}
The main limitation of the model is that it does not address changes in responses over time. The parallels between deep networks and the visual cortex are strongest in the context of short-term responses to brief stimuli, but responses in the brain develop more fully on longer timescales. For example, in the ventral stream, inferotemporal neurons become more selective for specific stimuli after a few tens of milliseconds \citep{Tamura2001}. In MT, neurons have temporal receptive fields with diverse biphasic profiles \citep{Perge2005}. Surround interactions evolve over tens of milliseconds \citep{Huang2007}. There is also a gradual evolution in the population response to objects' actual motion directions, versus the components of this motion that are orthogonal to local edges \citep{Pack2001,Smith2005}. On longer timescales, adapation-related effects develop over tens of seconds \citep{Kohn2004}, narrowing direction tuning and shifting preferred directions. Although our present focus is on training feedforward networks to emulate short-latency activity patterns, an important next step would be to address changes over time, perhaps through recurrent connections. 

There are also several additional details that are works in progress at the time of writing. For example, we have approximated distributions of pattern and component cells, and we have modelled various points in these distributions with weighted sums of tuning functions that are driven by optic flow fields with different numbers of pyramids. However, more work is needed to accurately model relationships between direction tuning width and correlations between neuron responses and idealized component and pattern responses. Further work is also needed to produce more accurate surround kernels, and to further compare our model with that of \citet{Baker2016}.   

\subsection{Empirical models vs. neural recordings for neural-representation learning}
A previous effort to train deep networks to approximate neural activity \citep{Yamins2014} may not have used a sufficiently large neural dataset. Suitably large chronic datasets are emerging \citep[e.g.][]{minderer2012chronic,obien2015revealing}. However, it is unlikely that we will soon have chronic recordings of many neurons in sulci of the intact cortex. Empirical models based on the electrophysiology literature can fill this gap. They require caution due to inevitable departures from real neural responses. However (as discussed in the Introduction) empirical models also seem to have certain advantages over real data. One is that the model properties can be modulated, allowing systematic investigation of a deep network's ability to reproduce various response features, and also allowing investigation of the influence of individual response features on task performance. Second, empirical models allow specification of an attention field at run-time. This should allow generation of attention-modulated activity labels that are consistent with the attention focus of the network, rather than the (perhaps different and/or unknown) attention focus of the animal.  

\subsection{Parallels between deep networks and the visual cortex}
\citet{Goodfellow-et-al-2016-Book} review some of the structural parallels between deep convolutional neural networks (CNNs) and the visual cortex. Among the similarities, both have feature hierarchies that develop through a sequence of linear combinations, nonlinearities, and pooling operations; both have sparse and spatially localized connections; and both represent features that are generally similar (in the cortex) or identical (in CNNs) across the visual scene. 
There are more subtle parallels as well. For example, divisive normalization is ubiquitous in the cortex \citep{Carandini2011} and similar operations can improve performance in deep networks \citep{Krizhevsky2012}. As another example, Poisson spike variability is ubiquitous in the cortex, and is related to Dropout \citep{Srivastava2014}, in that both consist of high response variability across repeated presentations of the same stimulus (although the statistics are somewhat different).  

Early layers of CNNs often converge on Gabor-like kernels \citep[e.g.][]{Zeiler2014}, which resemble receptive fields in the primary visual cortex (V1). Interestingly, splitting early layers and restricting connections between them has led to separation of edge-detecting and colour-detecting units \citep{Krizhevsky2012}, analogous to interstripes and blobs in V1. Similarities have also been reported between V2 and related networks \citep{Lee2007}. 

Recently, it was shown \citep{Yamins2014,Cadieu2014,Khaligh-Razavi2014,Hong2016} that internal representations of deep networks trained for object classification have strong parallels with representations in the ventral visual stream, which is the main path for object vision in the primate. Relationships between CNNs and the dorsal visual stream have been less studied, but \citet{Guclu2016} found that action-recognition CNNs were predictive of function magnetic resonance imaging data from the dorsal stream. 

There are also many differences between standard deep networks and the visual cortex, including learning mechanisms, the predominance of feedback and lateral connections in the latter, and the far greater complexity of neurons than CNN units. 
Furthermore, even in core object recognition \citep{Cadieu2014}, a feedforward process in which deep networks have much in common with the cortex, the representations are clearly different. Also, comparisons in the literature have not yet addressed a number of response properties, including responses to occlusion and clutter in complex scenes. Our preliminary unpublished observations suggest however that responses to clutter and occlusion are fairly similar in object-classification CNNs and IT. 

In general, the similarities suggest that further work to align CNNs with the primate visual cortex may lead to networks that approximate the visual cortex very closely. 



\section{Conclusion}
We have presented a detailed, pixel-computable, empirical model of activity in the primate middle temporal area (MT). This model closely approximates many tuning properties of MT neurons, and statistical distributions of many response parameters. The model also approximates an MT response phenomenon that has not previously been addressed in models, namely local motion integration. We hope to use the model as an infinite and customizable (but approximate) source of labels for training deep networks to approximate MT activity. 

\section*{Acknowledgements} 
This work was supported by Mitacs and CrossWing Inc. 

\bibliography{empirical-mt}
\bibliographystyle{apa}

\end{document}